\documentclass[a4paper,prd,showpacs,noshowkeys,preprint]{revtex4}
\usepackage[latin1]{inputenc}
\usepackage{amsmath}
\usepackage{amssymb}
\usepackage{mathrsfs}

\providecommand{\bp}{{\bf p}}
\providecommand{\bk}{{\bf k}}
\providecommand{\bq}{{\bf q}}
\providecommand{\bx}{{\bf x}}
\providecommand{\by}{{\bf y}}
\providecommand{\Dp}[1]{{d^3\!{#1}}\hs{1pt}}
\providecommand{\fpi}{(2\pi)^{\frac{3}{2}}}
\def\sp{{\cdot}}
\providecommand{\meio}{\mt{\frac{1}{2}}}
\providecommand{\fp}[1]{\frac{e^{#1}}{\fpi}}

\providecommand{\inc}{\mt{(-)}}
\providecommand{\oi}{\mt{(\pm)}}

\providecommand{\eq}[1]{\begin{equation} #1 \end{equation}}
\providecommand{\eqarr}[1]{\begin{eqnarray} #1 \end{eqnarray}}
\providecommand{\hs}[1]{\hspace{#1}}
\providecommand{\ms}[1]{\mbox{\small $#1$}}
\providecommand{\mss}[1]{\mbox{\scriptsize $#1$}}
\providecommand{\mt}[1]{\mbox{\tiny $#1$}}
\providecommand{\mfn}[1]{\mbox{\footnotesize $#1$}}
\providecommand{\tp}{\mss{\mathsf{T}}}
\providecommand{\bs}[1]{\boldsymbol{#1}}
\providecommand{\ket}[1]{\vert #1 \rangle}
\providecommand{\bra}[1]{\langle #1 \vert}
\providecommand{\braket}[2]{\langle #1\vert #2 \rangle}
\begin{document}
\title{Flavor mixing in a Lee-type model}
\author{C.~C.~Nishi\,$^{\rm(ab)}$}
\email{ccnishi@ifi.unicamp.br}
\author{M.~M.~Guzzo\,$^{\rm(a)}$}
\email{guzzo@ifi.unicamp.br}
\affiliation{$^{\rm(a)}$
Instituto de Física ``Gleb Wataghin''\\
Universidade Estadual de Campinas, Unicamp\\
13083-970, Campinas, SP, Brasil
}
\affiliation{$^{\rm(b)}$
Instituto de Física Teórica,
UNESP -- São Paulo State University\\
Rua Pamplona, 145,
01405-900 -- São Paulo, Brasil
}

\begin{abstract}
An exactly solvable Quantum Field Theory (QFT) model of Lee-type is
constructed to study how neutrino flavor eigenstates are created through
interactions and how the localization properties of neutrinos follows from
the parent particle that decays. The two-particle states formed by the neutrino
and the accompanying charged lepton can be calculated exactly as well as their
creation probabilities.
We can show that the coherent creation of neutrino flavor eigenstates follows
from the common negligible contribution of neutrino masses to their creation
probabilities. On the other hand, it is shown that it is not possible to
associate a well defined ``flavor'' to coherent superpositions of charged leptons.
\end{abstract}
\pacs{14.60.Pq, 13.15.+g, 11.10.-z}
\maketitle
\section{Introduction}
\label{sec:intro}

In recent times the neutrino oscillation phenomenon was established as an
example where the coherent creation of a superposition of mass eigenstates,
which is called a ``flavor'' eigenstate, tagged by the accompanying charged
lepton, leads to an interference among those mass eigenstates inducing the
flavor oscillation. We will choose to refer to those specific superpositions of
mass eigenstates as neutrino flavor states because there is no well
defined ``flavor'' charge operator to which we can associate flavor quantum
numbers. Moreover, we will try to show that the concept of flavor for neutrinos
is an approximate notion.

In this article a Lee-type model is deviced to investigate the neutrino flavor
creation problem. 
The Lee model\,\cite{lee,kallen-pauli} is an example of a simple quantum field
theory (QFT) which is exactly soluble but it still requires wave
function and charge renormalization to make the model meaningful.
Such features allied to strong conservation laws enable us to solve the model
completely for each sector invariant by the conservation law.
Another unique feature of this model, which will be explored here, is that it
permits the study of unstable particles in an exact way\;\cite{glaser-kallen}.
The modifications performed here consist on replacing the fields $N,\theta, V$
by $l_i,\nu_j,\pi$, where the unstable fermion $V$ in the original model is
replaced by the scalar boson $\pi$ and other fields are replaced accordingly.
Thus this model will be used to mimic the decay $\pi^{\pm}\rightarrow
l_\alpha^{\pm}\nu_\alpha$ where $\alpha=e,\mu$.

The decay process $\pi\rightarrow l_i\nu_j$, $i,j=1,2$, has four kinematical
decay channels available for two neutrino families. The channel $\pi\rightarrow
l_1\nu$ ($l_1=e$), with arbitrary neutrino content, is, however, very
suppressed with respect to the dominant $\pi\rightarrow l_2\nu$ ($l_2=\mu$)
because of helicity suppression in weak decays. The model deviced here does not
account for this helicity suppression, giving comparable branching ratios for
the production of both charged leptons. However, comparable branching ratios for
different charged leptons accompanied by neutrinos exist in nature\,\cite{pdg}:
(a) $K^0_L\rightarrow\pi^{\pm}e^{\mp}\nu_e$ (40.53\%) and
$K^0_L\rightarrow\pi^{\pm}\mu^{\mp}\nu_\mu$ (27.02\%), (b) $K^+\rightarrow
\pi^0e^+\nu_e$ (4.98\%) and $K^+\rightarrow \pi^0\mu^+\nu_\mu$
(3.32\%), and (c) $W^+\rightarrow l_\alpha\nu_\alpha$. The last process is also
effectively a two body decay kinematically analogous to the pion decay.

In the aforementioned processes we could in principle observe the interference
among neutrinos produced jointly with different charged leptons such as
$\pi\rightarrow l_1\nu_1$ and $\pi\rightarrow l_2\nu_1$. Why is this kind of
interference not observed? We will try to address this question within this
Lee-type model keeping in mind a fictitious pion decay without helicity 
suppression, with comparable branching ratios to decay into different charged
leptons. Previous discussions about the possibility of charged lepton flavor
oscillations exist in the literature\,\cite{akhmedov,dolgov} but we should
remark that the question raised previously is more general in the sense that we
could observe the interference of the channels by observing the neutrinos and
not the charged leptons. On the other hand, there is also the interesting
possibility that neutrino oscillations could be suppressed by small momentum or
energy uncertainties\,\cite{kayser:81,akhmedov:mossbauer}.

A throughout study of the mixing phenomenon, either in the bosonic or
fermionic case, was already performed in Quantum Field Theory (QFT) by Blasone
and Vitiello (BV)\,\cite{BV}. Despite of that, the role played by the
interaction which generates the mixing was not completely carried out.
Although an analysis of flavor states emerging from pion decay is shown in
Ref.\,\onlinecite{BV:pi}.
In an almost free QFT, BV analyzes the effects of mixing and seek for what
should be the flavor states for neutrinos. What emerges from that
study is the possibility to define two types of unitarily inequivalent states
constructed from two distinct vacua: a usual vacuum and a mixed vacuum. They use
the mixed vacuum and obtain a different neutrino oscillation phenomenon. The
choice of the vacuum, however, does not seem to be unique. The study of a Lee
type model presents two advantages with respect to the BV approach: (i) the
interaction responsible by flavor mixing can be taken into account exactly and
(ii) the free states are unitarily equivalent to exact states of the total
Hamiltonian. The point (ii) means we do not have to choose between two unitarily
inequivalent set of states but at the same time we can not study in this type of
models the consequences of such phenomenon, which might be present in more
realistic QFTs.

Another distinctive approach to treat neutrino oscillations in a rigorous
manner was extensively reviewed in Ref.\,\onlinecite{beuthe}, where the so
called external wave packet (EWP) approach was analyzed. Although, the first QFT
treatment of neutrino oscillations was given in Ref.\,\onlinecite{kobzarev}. In the
EWP approach, the nonobservability of neutrinos through direct detection was
translated into the formalism by treating them as internal Feynman propagators
connecting the creation and detection processes. Some unclear aspects of such
approach was clarified in Ref.\,\onlinecite{ccn:no12}. One of them concerned the
possibility that some nonphysical contributions, such as below threshold neutrino
detection, could be present in the formalism and they have to be removed by
hand\,\cite{ccn:no12}. Such approach could account for the localization aspects of
the creation and detection processes\,\cite{kiersWeiss,rich:no,grimus} but the
meaning of what would be the intermediate neutrino flavor states could not be
investigated. Another attempt to define neutrino flavor states in QFT can be found in
Ref.\,\onlinecite{giunti:qft}. One can also find attempts to describe neutrino
oscillations in Relativistic Quantum Mechanics\,\cite{bernardini,ccn:no12}.

The two goals of this article consist in providing an exactly solvable model to
study (a) how neutrinos are created through interactions and (b) how the
localization of neutrinos follows from the parent pion properties. As is usually
adopted in rough estimates the wave packet size of the daughter particles is
considered to be $1/\Gamma$\,\cite{akhmedov,kayser:81}, where $\Gamma$ is the
decay width of the parent particle. This model provides exactly such relation.
We can also calculate the two particle state composed by the charged lepton and
the neutrino that is created from pion decay. This calculation is an attempt to
properly define a flavor state for neutrinos. We will see that the coherent creation
of neutrino flavor states is a consequence of the common negligible kinematical
contribution of the different neutrino mass eigenstates to their creation
probabilities (amplitudes). We will also study the meaning of defining a different
``flavor'' state to charged leptons as the coherent superposition that
accompanies a neutrino mass eigenstate such as $\nu_1$, for instance. We will
conclude that the concept of ``flavor'' associated to such superposition states can
not be properly defined.

The outline of the article is as follows: in Sec.\,\ref{sec:model} we
introduce the model and calculate the relevant eigenstates of the total
Hamiltonian. In Sec.\,\ref{sec:flavor} we study how the concept of
neutrino flavor emerges from the calculations and derive the localization
properties of the decaying states. The discussions are made in
Sec.\,\ref{sec:discussion} and some detailed or additional calculations are
shown in the appendices.

\section{The model}
\label{sec:model}

The free Hamiltonian of the model is defined by
\eqarr{
\label{H0}
H_0&\equiv&\mu\int\Dp{k}A^\dag(\bk)A(\bk)
+ M_i\int\Dp{p}b^\dag_i(\bp)b_i(\bp)\cr
&&
+\int\Dp{p}E_j(\bp)a^\dag_j(\bp)a_j(\bp)
\,,
}
where
\eq{
\label{E:nu}
E_j(\bp)=\sqrt{\bp^2+m^2_j}\,,
}
and $A,b_i$ and $a_j$ are respectively the annihilation operators of the
fields $\pi,l_i$ and $\nu_j$ modelling (unrealistically) the decay
$\pi\rightarrow l_i\nu_j$. The summation over repeated indices are implicit.
We will restrict ourserves to two families
which means $l_i=l_1,l_2$ ($l_1\equiv e,l_2\equiv\mu$) and $\nu_j=\nu_1,\nu_2$
correspond to the two neutrino mass eigenstates. Extension to three neutrino
families is straightforward.
The fixed bare energies of the fields $\pi$ and $l_i$ corresponding to their
masses $\mu,M_i$ mean this model treats the static and recoilless limit of
$\pi$ and $l_i$. The momentum though is conserved.
The neutrinos, on the other hand, obey the relativistic energy
dispersion relation \eqref{E:nu}.
The creation and annihilation operators satisfy
\eqarr{
[A(\bk),A^\dag(\bk')]&=&\delta^3(\bk-\bk')\\
\{b_i(\bp),b_j^\dag(\bp')\}
&=&\delta_{ij}\delta^3(\bp-\bp')\\
\{a_i(\bp),a_j^\dag(\bp')\}
&=&\delta_{ij}\delta^3(\bp-\bp')
}
The spin indices are suppressed.

We define the Fourier transform of the fields to be
\eqarr{
l_i(\bx)&\equiv &
\int\Dp{p}b_i(\bp,r)u_0^r\fp{i\bp\sp\bx}
\,\\
\pi(\bx)&\equiv &
\int\Dp{k}A(\bk)\fp{i\bk\sp\bx}
\,\\
\label{nu:dp}
\nu_i(\bx)&\equiv &
\int\frac{\Dp{p}}{\sqrt{2E_i}}\big[
a_i(\bp,r)u_i^r(\bp)\fp{i\bp\sp\bx}+
a_i^\dag(\bp,r)\eta_C C \overline{u_i^r}^{\tp}(\bp)
\fp{-i\bp\sp\bx}\big]\,,
}
where $u_i^r(\bp)$, $r=1,2$, are Dirac spinors normalized as
$u_i^{r\dag}(\bp)u_i^s(\bp)=2E_i(\bp)\delta_{rs}$ while $u_0^r$, $r=1,2$, are 
the corresponding spinors for fermions at rest properly normalized to
$u_0^{r\dag}u_0^s=\delta_{rs}$.
Expression \eqref{nu:dp} corresponds to the usual Majorana type fermion
expansion where $\eta_C$ is a phase factor appearing in the charge conjugation
transformation.

The interaction Hamiltonian is chosen to be
\eqarr{
H_1&=&
g_0
\int \Dp{x}\Dp{y}\big[
l_i^\dag(\bx)U_{ij}\nu_j^{\inc}(\bx)\pi(\by)f(\bx-\by)+h.c.
\big]
\\&=&
g_0U_{ij}
\int\Dp{p}\Dp{p'}
b_i^\dag(\bp,r)\eta^{rs}_j(\bp')a_j^\dag(\bp',s)A(\bp+\bp')
+h.c.,
}
where $\{U_{ij}\}$ corresponds to the mixing matrix, $\nu^{\inc}_j$ refers to
the creation part of expansion \eqref{nu:dp} and 
\eq{
\eta^{rs}_j(\bp')=
\frac{1}{\fpi}
\frac{f(\bp')}{\sqrt{2E_j(\bp')}}
\eta_Cu_0^{r\dag}C\overline{u_j^s}^{\tp}(\bp')
\,.
}
Using the Dirac representation for gamma matrices, $C=i\gamma_2\gamma_0$,
$u_0^r=e_r$\,\cite{endnote0} and
$u_j^s(\bp)=\frac{E_j\gamma_0-\bp\sp\bs{\gamma}+m_j}{\sqrt{m_j+E_j}}u_0^s$ we
obtain
\eq{
\tilde{\eta}_{rs}(\bp)\equiv
\eta_Cu_0^{r\dag}C\overline{u_j^s}^{\tp}(\bp)=
\frac{-i\eta_C(\bs{\sigma}\sp\bp\sigma_2)_{rs}}{\sqrt{m_j+E_j}}\,.
}
The function $f$ represents a ``form factor'' necessary to regularize the
expressions in the Lee model\,\cite{kallen-pauli}. For consistency reasons to be
discussed later it also has to be smooth, analytic, approximately flat for
energies much lower than the cutoff scale, but falling off not so rapidly above
the cutoff scale. In addition, the Lee model may violate
unitarity if the coupling constant is above a critical
value\,\cite{kallen-pauli}. We assume the coupling constant is below the
critical value thus avoiding such situation. Nevertheless, the case of
an unstable parent particle can be properly described in the unitary
regime\,\cite{glaser-kallen}.

There are two charges $Q_1,Q_2$ that are conserved and within those sectors of
fixed charges we can find the explicit eigenstates of the whole Hamiltonian
\eq{
H=H_0+H_1\,.
}
The two charges are
\eqarr{
Q_1&=& \int\Dp{k}A^{\dag}(\bk)A(\bk)+
\sum_i\int\Dp{p}b_i^{\dag}(\bp)b_i(\bp)\,,
\\
Q_2&=& \sum_i\int\Dp{p}[b_i^{\dag}(\bp)b_i(\bp)
-a_i^{\dag}(\bp)a_i(\bp)]
\,.
}

In the sector containing one $\pi$ [$(Q_1,Q_2)=(1,0)$] and one pair of
$l_i,\nu_j$ [$(Q_1,Q_2)=(1,0)$] we can calculate the exact eigenstates of $H$ in
terms of the following eigenstates of $H_0$
\eqarr{
\label{pi:0}
\ket{\pi}_0&\equiv&\int\Dp{k}\psi_\pi(\bk)\ket{\pi(\bk)}_0\,,\\
\label{linuj:0}
\ket{l_i(r),\nu_j(\bp',s)}_0&\equiv&\int\Dp{p}\psi_\pi(\bp+\bp')
\ket{l_i(\bp,r),\nu_j(\bp',s)}_0 
\,,
}
where
\eqarr{
\ket{\pi(\bk)}_0&\equiv&A^\dag(\bk)\ket{0}\,,\\
\ket{l_i(\bp,r),\nu_j(\bp',s)}_0&\equiv&
b_i^\dag(\bp,r)a_j^\dag(\bp',s)\ket{0}
\,,
}
and $\ket{0}$ is the vacuum state of $H_0$ and of $H$ as well\,\cite{endnote1}.
The function $\psi_\pi$ is an arbitrary function controlling the pion momentum
distribution.
The subscript $0$ indicates the states correspond to eigenstates of $H_0$, i.e.,
bare states. Then we find the eigenvalue equations
\eqarr{
H_1\ket{\pi}_0
&=& 
g_0U_{ij}\int\Dp{p'}\eta^{rs}_j(\bp')
\ket{l_i(r),\nu_j(\bp',s)}_0
\,,\\
H_1\ket{l_i(r),\nu_j(\bp',s)}_0
&=&
g_0U^*_{ij}\eta^{rs*}_j(\bp')\ket{\pi}_0
\,.
}

The eigenstate of $H$ with energy (mass) $E_\pi$ corresponding to a dressed
state of $\ket{\pi}_0$ can be found if we discover the function $\chi$ in
\eq{
\label{pi:ket}
\ket{\pi}=\ket{\pi}_0+
\sum_{ij,rs}\int\Dp{p'}\chi_{ij}^{rs}(\bp')\ket{l_i(r),\nu_j(\bp',s)}_0
\,.
}
From the eigenvalue equation
\eq{
H\ket{\pi}=E_\pi\ket{\pi}\,,
}
we obtain
\eq{
\chi_{ij}^{rs}(\bp')= 
-\frac{g_0U_{ij}\eta^{rs}_j(\bp')}
{M_i+E_j(\bp')-E_\pi}
\,
}
and $h_0(E_\pi)=0$ where
\eq{
\label{h0=0}
h_0(E_\pi)\equiv E_\pi-\mu+\phi_0(E_\pi)\,.
}
The function $\phi_0$ is defined as 
\eq{
\label{phi0:def}
\phi_0(E)\equiv
g^2_0
\sum_{\underset{\mfn{rs}}{ij}}
\int\Dp{p'}\frac{|\eta^{rs}_j(\bp')|^2|U_{ij}|^2}
{M_i+E_j'-E}\,.
}
We assume there is only one root for Eq.\,\eqref{h0=0} which defines the pion
mass $M_\pi$, i.e., the energy of the dressed state $\ket{\pi}$. 
Therefore we could have written the bare pion mass $\mu$ as
\eq{
\mu=M_\pi+\delta\mu\,,
}
where
\eq{
\delta\mu=\phi_0(M_\pi)\,
}
is the mass counterterm.

We can simplify
\eq{
\label{phi0:phitilde}
\phi_0(E)=
\gamma_0\sum_{ij}|U_{ij}|^2\tilde{\phi}_j(E-M_i)
\,,
}
where
\eq{
\label{g0}
\gamma_0\equiv \frac{g^2_0}{2\pi^2}\,
}
and
\eq{
\label{phitilde}
\tilde{\phi}_j(x)\equiv
\int_{m_j}^{\infty}dE\sqrt{E^2-m^2_j}(E-m_j)
\frac{f^2(E)}{E-x}
\,.
}
The intermediate steps to reach Eq.\,\eqref{phi0:phitilde} can be found
in appendix \ref{ap:phi0}. We should emphasize, however, that the principal
value is understood in Eq.\,\eqref{phitilde} if $x>m_j$\,\cite{glaser-kallen}.

To have an unstable pion we should have\,\cite{glaser-kallen}
\eq{
\label{unstable:h0}
h_0(M_1+m_1)<0
}
for $M_1=\min(M_i)$ and $m_1=\min(m_j)$. This condition implies that 
\eq{
\label{unstable:Mpi}
M_\pi> M_1+m_1\,.
}
These conclusions follow from $\phi_0(-\infty)=0-\epsilon$,
$d\phi_0(x)/dx>0$ for $x<m_1+M_1$ and $\phi_0(x)>0$ for
 $x\rightarrow m_1+M_1-\epsilon$, which assures $h_0(x)$ is
a monotonically increasing function for $x< M_1+m_1$. 
Equation \eqref{unstable:h0} guarantees that the root of $h_0(E)=0$, the pion
mass, is above the threshold $M_1+m_1$, leading thus to
Eq.\,\eqref{unstable:Mpi}.
It can be also seen from Eq.\,\eqref{phitilde} that for $x>M_1+m_1$, $\phi_0(x)$
may continue increasing but it starts to decrease when $x$ is greater than the
cutoff scale of $f$. In addition we will assume all the decaying channels are
open, i.e., $M_\pi>m_2+M_2$. Intermediate cases can be handled with appropriate
care.

If condition \eqref{unstable:Mpi} is satisfied, Eq.\,\eqref{pi:ket} does not
define a meaningful expansion although the norm is finite
\eq{
\label{pi:norm}
\braket{\pi}{\pi}=h'_0(M_\pi)=1+\phi'_0(M_\pi)\,,
}
with $h'_0(x)=dh_0(x)/dx$.
For stable pion, the norm \eqref{pi:norm} is used to (re)normalize the pion
state since the expression in Eq.\,\eqref{pi:norm} diverges when there
is no cutoff, i.e., $f(E)=1$.
It means there is no stable dressed pion state and the only stable eigenstates
of $H$ in this sector (one $\pi$ or one pair of $l_i,\nu_j$) are the scattering
states containing one $l_i$ and one $\nu_j$. These states complete the Hilbert
space in this sector.
This is proved in appendix \ref{linuj:completo}.

The scattering states, eigenstates of $H$, can be calculated from the
expansion
\eqarr{
\label{linuj:d}
\ket{l_i(r),\nu_j(\bp,s)}&=&
\ket{l_i(r),\nu_j(\bp,s)}_0
+\int\Dp{p'}\alpha^{rs,r's'}_{ij,i'j'}(\bp,\bp')
\ket{l_{i'}(r'),\nu_{j'}(\bp',s')}_0
\cr&&~
+~Z_2^{\frac{1}{2}}\beta^{rs}_{ij}(\bp)\ket{\pi}_0
\,,
}
if we find the functions $\alpha$ and $Z_2^{\meio}\beta$.
The eigenvalue equation
\eq{
H\ket{l_i(r),\nu_j(\bp,s)}=
\big(M_i+E_j(\bp)\big)\ket{l_i(r),\nu_j(\bp,s)}\,,
}
leads to
\eqarr{
\label{alpha-beta}
(M_i+E_j(\bp)-M_{i'}-E_{j'}(\bp'))\alpha^{rs,r's'}_{ij,i'j'}(\bp,\bp')
&=& g_0Z_2^{\frac{1}{2}}
\beta^{rs}_{ij}(\bp)\eta^{r's'}_{j'}(\bp')U_{i'j'}\,,\\
\label{beta-alpha}
(M_i+E_j(\bp)-M_\pi-\delta\mu)Z_2^{\frac{1}{2}}\beta^{rs}_{ij}(\bp)
&=&
g_0\eta^{rs*}_j(\bp)U^*_{ij}+
g_0\int\Dp{p'}\alpha^{rs,r's'}_{ij,i'j'}\eta^{r's'*}_{j'}U^*_{i'j'}
\,.~~
}
Inverting Eq.\,\eqref{alpha-beta} using the incoming (+) or outgoing (-) wave
boundary conditions we obtain
\eq{
\alpha^{rs,r's'\oi}_{ij,i'j'}(\bp,\bp')
=
\frac{g_0Z_2^{\frac{1}{2}}\beta^{rs\oi}_{ij}(\bp)\eta^{r's'}_{j'}(\bp')U_{i'j'}}
{M_i-M_{i'}+E_j(\bp)-E_{j'}(\bp')\pm i\epsilon}
\,.
}
Equation \eqref{beta-alpha} yields
\eq{
Z_2^{\frac{1}{2}}\beta^{rs\oi}_{ij}(\bp)=
\frac{g_0\eta^{rs*}_j(\bp)U^*_{ij}}{h_0(\ms{M_i+E_j(\bp)\pm i\epsilon})}
\,.
}
Notice that the domain of the function $h_0$ was extended to the complex
numbers, hence satisfying the property
\eq{
\label{h0(x+i)}
h_0(E\pm i\epsilon)=
h_0(E)\pm \frac{i}{2}\Gamma_0(E)
\,,
}
where
\eqarr{
\label{Gamma:0}
\Gamma_0(E)&\equiv&
\gamma_0\sum_{ij}
|U_{ij}|^2\tilde{\Gamma}_j(E-M_i)
\,,
\\
\label{Gamma:j}
\tilde{\Gamma}_j(x)&\equiv&
2\pi\theta(x-m_j)(x-m_j)\sqrt{x^2-m^2_j}f^2(x)
\,.
}
Therefore
\eq{
\alpha^{rs,r's'\oi}_{ij,i'j'}(\bp,\bp')=
\frac{g^2_0U^*_{ij}U_{i'j'}}{h_0(\ms{M_i+E_j(\bp)\pm i\epsilon})}
\frac{\eta^{rs*}_j(\bp)\eta^{r's'}_{j'}(\bp')}
{M_i+E_j(\bp)-M_{i'}-E_{j'}(\bp')\pm i\epsilon}
\,.
}

Despite the divergence in Eq.\,\eqref{pi:norm}, we can still define the
approximate pion state
\eq{
\label{pi:lambda:0}
\ket{\pi_\lambda}=C\big[
\ket{\pi}_0 - 
g_0U_{ij}\int\Dp{p'}\frac{\eta^{rs}_{j}(\bp')}{M_i+E_j(\bp')-M_\pi-i\lambda}
\ket{l_i(r),\nu_j(\bp',s)}_0
\big]\,,
}
with $M_\pi$ being the root of Eq.\,\eqref{h0=0}. Such state can
be properly normalized if $\lambda\neq 0$. That this state is not an exact
eigenstate of $H$ is proved in appendix \ref{ap:pi:approx}.
If we expand the state \eqref{pi:lambda:0} in terms of the scattering states
\eqref{linuj:d}, we can see how this state can be seen as an approximate
ressonant state corresponding to the pion that decays.

We calculate, for outgoing states, with superscript $(-)$ suppressed,
\eqarr{
\braket{l_i(r),\nu_j(\bp,s)}{\pi}_0&=&
Z_2^{\frac{1}{2}}\beta^{rs*}_{ij}(\bp)
\\
\braket{l_i(r),\nu_j(\bp,s)}{l_{i'}(r),\nu_{j'}(\bp',s)}_0&=&
\delta^{3}(\bp-\bp')\delta_{rr'}\delta_{ss'}\delta_{ii'}\delta_{jj'}\,,
}
which yield
\eqarr{
\ket{\pi_\lambda}&=&C\sum_{\overset{\mfn{rs}}{ij}}
\int\Dp{p}\ket{l_i(r),\nu_j(\bp,s)}
\frac{g_0U_{ij}\eta^{rs}_{j}(\bp)}
{h_0(M_i+E_j+i\epsilon)(M_i+E_j-M_\pi-i\lambda)}
\cr&&\hs{5em}\times
\big[-i\lambda+\phi_0(M_\pi)-\phi_0(M_\pi+i\lambda)\big]
\,.
}

For small $\lambda$, i.e., $|\lambda|\ll \Gamma_0(M_\pi)/2\phi'_0(M_\pi)$, we
can approximate
\eqarr{
\label{pi:lambda}
\ket{\pi_\lambda}&\approx &
-i\lambda \Big[1+\frac{\Gamma(M_\pi)}{2|\lambda|}\Big]^{\frac{1}{2}}
\sum_{\overset{\mfn{rs}}{ij}}
\int\Dp{p}\ket{l_i(r),\nu_j(\bp,s)}
\cr&&\hs{5em}\times
\frac{gU_{ij} \eta^{rs}_{j}(\bp)}
{h(M_i+E_j+i\epsilon)(M_i+E_j-M_\pi-i\lambda)}
\,,
}
We have defined the renormalized coupling constant
\eq{
\label{g0->g}
g\equiv Z_2^{\frac{1}{2}}g_0,~~\text{or }
~\gamma\equiv Z_2\gamma_0\,,~~
}
which defines the ``renormalized'' functions $\Gamma,\phi$ obtained by replacing
$\gamma_0$ by $\gamma$ in $\Gamma_0,\phi_0$, thus multiplicatively, while
\eq{
h(E)\equiv Z_2h_0(E)\,.
}
The renormalization constant is
\eq{
Z_2^{-1}=1+\phi'_0(M_\pi)=\frac{1}{1-\phi'(M_\pi)}\,.
}

Since the scattering states $\ket{l_i(r),\nu_j(\bp,s)}$ are
eigenstates of the total Hamiltonian $H$ with energy $E=M_i+E_j(\bp)$, we can
easily compute 
\eqarr{
\label{pipi:t}
\braket{\pi_\lambda}{\pi_\lambda(t)}&=&
\lambda^2\Big[1+\frac{\Gamma(M_\pi)}{2|\lambda|}\Big]
\int_{m_1+M_1}^{\infty}\!\!\!dE\,\frac{\Gamma(E)}{2\pi}
\cr&&\times
\label{e-iEt}
\frac{e^{-iEt}}
{\displaystyle\big[(E-M_\pi+\phi^{\mt{(2)}}(E))^2+\frac{\Gamma^2(E)}{4}\big]
\big[(E-M_\pi)^2+\lambda^2\big]}
\,,~~
}
where $\ket{\pi_\lambda(t)}=e^{-iHt}\ket{\pi_\lambda}$,
$\phi^{\mt{(2)}}(E)\equiv\phi(E)-\phi(M_\pi)-\phi'(M_\pi)(E-M_\pi)$ and
Eq.\,\eqref{Gamma:0} were used. 

If the functions $\Gamma(E),\phi(E)$ vary more slowly than $E$ near $E=M_\pi$
we can approximate $\Gamma(E)\approx \Gamma(M_\pi)$ and if $M_\pi\gg M_i+m_j$
we can replace the lower limit of the integral in Eq.\,\eqref{e-iEt} by
$-\infty$:
\eqarr{
\label{pipi(t)}
\braket{\pi_\lambda}{\pi_\lambda(t)}&\approx&
e^{-iM_\pi t}\Big[1+\frac{\Gamma_\pi}{2|\lambda|}\Big]^{-1}
\Big[e^{-\Gamma_\pi t/2}-\frac{\Gamma_\pi}{2|\lambda|}e^{-|\lambda|t}\Big]
\\
\label{e-Gt}
&\stackrel{|\lambda|\gg \Gamma_\pi}{\approx}&e^{-i(M_\pi-i\Gamma_\pi/2 )t}
\,,
}
where
\eq{
\Gamma_\pi\equiv \Gamma(M_\pi)\,.
}
Equation \eqref{e-Gt} expresses the usual exponential decay law with decay
rate $\Gamma_\pi$. This result is an indication that the approximate state
$\ket{\pi_\lambda}$ in Eq.\,\eqref{pi:lambda:0} represents appropriately a
ressonant pion state that decays into $\ket{l_i(r),\nu_j(\bp,s)}$ states
independently of the arbitrary parameter $\lambda$. If we had calculated
Eq.\,\eqref{pipi(t)} for a state analogous to Eq.\,\eqref{pi:lambda:0} but with
approximate energy very different from $M_\pi$ we would obtain a decaying
amplitude in Eq.\,\eqref{e-Gt} with decay rate primarily given by
$|\lambda|\gg\Gamma_\pi$ (see proper discussion in
Ref.\,\onlinecite{glaser-kallen}).
It is important to remark that $|\lambda|$ should be also sufficiently
small to allow Eq.\,\eqref{h0(x+i)} when $\epsilon$ is replaced by $\lambda$.
Combining $|\lambda|\gg \Gamma_\pi$ and $|\lambda|\ll
\Gamma_\pi/2\phi'(M_\pi)$, it is necessary that $\phi'(M_\pi)\ll 1$.

The emission spectrum of $l_i,\nu_j$ is given by 
\eq{
|\braket{l_i(r),\nu_j(\bp,s)}{\pi_\lambda(t)}|^2=|C^{rs}_{ij}(\bp,
\lambda)|^2\,,
}
where
\eq{
C^{rs}_{ij}(\bp,\lambda)=
-i\lambda \Big[1+\frac{\Gamma_\pi}{2|\lambda|}\Big]^{\frac{1}{2}}
\frac{gU_{ij} \eta^{rs}_{j}(\bp)}
{h(M_i+E_j+i\epsilon)(M_i+E_j-M_\pi-i\lambda)}
}
is the expansion coefficient in Eq.\,\eqref{pi:lambda}. We can see
\eq{
\sum_{\overset{\mfn{rs}}{ij}}\int\Dp{p}
|C^{rs}_{ij}(\bp,\lambda)|^2=
\sum_{ij}\int_{m_1+M_1}^{\infty}dE\,W_{ij}(E-M_i,\lambda)
=1\,.
}
The function $W_{ij}(E,\lambda)$ is the emission probability for $\pi\rightarrow
l_i\nu_j$, per unit energy, with neutrino energy $E$, summed over spins:
\eq{
W_{ij}(E-M_i,\lambda)=
\gamma \tilde{\Gamma}_j(E-M_i)|U_{ij}|^2
\frac{W_\lambda(E)}{\Gamma(E)}
\,.
}
The total emission probability is
\eq{
W_\lambda(E)
=
\frac{1}{2\pi}
\frac{\Gamma(E)}{h^2(E)+\Gamma^2(E)/4}
\frac{\lambda^2[1+\Gamma_\pi/2|\lambda|]}{(E-M_\pi)^2+\lambda^2}
}
which approximates the usual Breit-Wigner distribution in the
$\lambda\gg \Gamma_\pi$ limit, within the approximations $h(E)\approx E-M_\pi$
and $\Gamma(E)\approx \Gamma_\pi$,
\eq{
\label{W:approx}
W(E)\equiv \lim_{\lambda\rightarrow\infty}W_\lambda(E)=
\frac{1}{2\pi}\frac{\Gamma_\pi}{(E-M_\pi)^2+\Gamma^2_\pi/4}
\,.
}
From probability conservation we have
\eq{
\sum_{ij}W_{ij}(E-M_i,\lambda)=W_\lambda(E)\,.
}
The energy $E$ in $W_\lambda(E)$ refers to the total energy of the pair $l+\nu$.

We can see the pion state is a ressonant state by calculating the scattering
cross section of $l_i\nu_j\rightarrow l\nu$, irrespective of the final product
flavors and spins and averaged over initial spins:
\eq{
\frac{d\sigma}{d\Omega}(l_i\nu_j\rightarrow l\nu;E)=
\frac{|U_{ij}|^2}{(E-M_i)^2-m^2_j}
\gamma \tilde{\Gamma}_j(E-M_i)\,\frac{\pi}{4} W(E)
\,,
}
where $E=E_j+M_i$ corresponds to the total energy of the incident
particles. The kinematics is totally determined because the
energy of $l_i$ and $\pi$ are fixed. The difference between the initial and
final momenta (velocities) of neutrinos has to be taken into account to obtain
the appropriate cross section obeying
\eq{
\sum_{ij}\frac{d\sigma}{d\Omega}(ij\rightarrow l\nu;E)=
\frac{\pi}{4\bp^2}\Gamma(E)W(E)\Big|_{\bp^2=(E-M_i)^2-m^2_j}
\,.
}
The presence of $W(E)$ guarantees the resonance for $E=M_\pi$.

\section{Flavor (eigen)states}
\label{sec:flavor}

In the previous section we have seen, from the QFT point of view, it makes
more sense to calculate transition probabilities (amplitudes) with respect to
the state $\ket{l_i(r),\nu_j(\bp,s)}$ that has definite energy
$E_j(\bp)+M_i$ and thus are mass eigenstates. What wee see in the actual pion
decay, however, is not $\pi\rightarrow l_i\nu_j$, $i,j=1,2$, but $\pi\rightarrow
l_\alpha\nu_\alpha$, with a definite flavor $\alpha=e,\mu$. For the charged
leptons the terminology ``flavor'' is clear and coincides with the mass
eigenstates $l_i=l_\alpha$. For neutrinos what characterizes a flavor state
depends on the accompanying charged lepton. 

Let us define flavor states as
\eq{
\label{lnu:alpha}
\ket{l_\alpha(r),\nu_\beta(\bp,s)}\equiv
\delta_{\alpha i}U_{\beta j} \ket{l_i(r),\nu_j(\bp,s)}
\,,~~\alpha,\beta=e,\mu\,.
}
Let us then calculate
\eqarr{
|\braket{l_\alpha,\nu_\beta(\bp)}{\pi_\lambda(t)}|^2&=&
\sum_{rs}\Big|\sum_jU^*_{\beta j}C^{rs}_{\alpha j}(\bp)
e^{-i(M_\alpha+E_j(\bp))t}\Big|^2
\\&=&
\label{creationP:2}
g^2\sum_{rs}\Big|\sum_jU_{\alpha j} e^{-iE_jt}(U^\dag)_{j\beta}
\frac{\eta^{rs}_j(\bp)}{h(E_j+M_\alpha+i\epsilon)}\Big|^2
\,,
}
already in the $\lambda\gg\Gamma_\pi$ limit.
If we take the $m_j\rightarrow 0$ limit ($E_j\rightarrow E_\nu=|\bp|$) in the
terms except in the exponent we arrive at
\eq{
\label{pi->l,Unu}
|\braket{l_\alpha,\nu_\beta(\bp)}{\pi_\lambda(t)}|^2=
\mathcal{P}_{\alpha\beta}(t)
\frac{W_{\alpha}(E_\nu)}{4\pi|\bp|E_\nu}
\,,
}
where
\eq{
\mathcal{P}_{\alpha\beta}(t)=
\Big|\sum_jU_{\alpha j}e^{-iE_jt}U^*_{\beta j}\Big|^2\,,
}
and $W_{\alpha}(E)$ is $W_{\alpha j}(E)$ in the $m_j\rightarrow 0$ limit and with
$|U_{ij}|^2$ factored out.
We see the probability of finding the flavor state \eqref{lnu:alpha} from
pion decay decouples into the usual oscillation probability times the
emission probability $\pi\rightarrow l_\alpha\nu$ irrespective of the neutrino
flavor.

The concept of neutrino flavor states emerges from Eq.\,\eqref{pi->l,Unu}
taking $t=0$ (or $t\ll$ oscillation period), for which we obtain
\eq{
\mathcal{P}_{\alpha\beta}(t\approx 0)=\delta_{\alpha\beta}
\,.
}
Therefore only a neutrino of flavor $\alpha$ is produced jointly with the
charged lepton of flavor $\alpha$.
If we try to calculate the creation probability for the equally possible
superposition state, corresponding to a neutrino mass eigenstate accompanied by a
superposition state of charged leptons,
\eq{
\label{pi->Ul,nu}
|\sum_i U^*_{i\beta}\braket{l_i,\nu_j(\bp)}{\pi_\lambda(t)}|^2 \,,
}
we see it does not decouple and since the masses $M_i$ have
different kinematical contributions, there is no way to define a superposition state
of charged leptons associated to a neutrino mass eigenstate $\nu_1$ for instance.
Stated differently, the transition probability \eqref{pi->Ul,nu} is not
equal to $\delta_{\beta j}$ for $t=0$, not even approximately.

The definition of the neutrino flavor state in Eq.\,\eqref{lnu:alpha} was
justified only a posteriori as the state that accompanies a definite
charged lepton. In this model, however, we could, in principle, calculate the
more interesting quantity
\eq{
\label{psi(x,t)}
\braket{l_i(r),\nu_j(\bx,s)}{\pi_\lambda(t)}=
\Psi^{rs}_{ij}(\bx,t)e^{-iM_it}\,,
}
where
\eq{
\ket{l_i(r),\nu_j(\bx,s)}\equiv
\int\Dp{p}\fp{-i\bp\sp\bx}\ket{l_i(r),\nu_j(\bp,s)}\,.
}
This ``wave function'' squared ($|\Psi^{rs}_{ij}(\bx,t)|^2$) would give us the
probability density of finding the neutrino $\nu_j$ in the position $\bx$ at
time $t$ jointly with a charged lepton $l_i$. The exact information of
$\Psi^{rs}_{ij}(\bx,t)$ allow us to expand
\eq{
\label{pi:psi(x,t)}
\ket{\pi_\lambda(t)}=
\sum_{ij,rs}\int\Dp{x}\Psi^{rs}_{ij}(\bx,t)e^{-iM_it}\ket{l_i(r),\nu_j(\bx,s)}
\,.
}
Equation \eqref{pi:psi(x,t)} characterizes completely how the neutrinos and
charged leptons arise from pion decay. Although the spatial information of the
charged lepton is not determined or calculable in this model.

The exact Fourier transform necessary to compute Eq.\,\eqref{psi(x,t)} is very
difficult to be performed analytically but a rough approximate calculation can
be performed. We find, in the limit $|\lambda|\gg \Gamma_\pi$,
\eqarr{
\label{Psi:x,t}
\Psi^{rs}_{ij}(\bx,t)&=&
\int\Dp{p}e^{-iE_j(\bp)t}C^{rs}_{ij}(\bp)\fp{i\bp\sp\bx}
\,,\\&\approx&
\label{Psi:x,t:approx}
-\frac{e^{-i\bar{E}_j t}}{\sqrt{4\pi}\,r}
\frac{\gamma^{\meio}}{\sqrt{\bar{v}_j}}
U_{ij}\tilde{\Gamma}^{\meio}_j(\bar{E}_j)
\frac{(\bs{\sigma}\sp\hat{\bx}\sigma_2)_{rs}}{\sqrt{2}}
\Big[
\theta(\tau_-)e^{i\bar{p}_jr}e^{-\frac{\Gamma_\pi}{2}\tau_-}
+\theta(\tau_+)e^{-i\bar{p}_jr}e^{-\frac{\Gamma_\pi}{2}\tau_+}
\Big]
\,,~~
}
where 
\eq{
\tau_{\pm}\equiv t\pm r/\bar{v}_j
\,,
}
$r=|\bx|$ and the bar in $\bar{v}_j,\bar{p}_j$ and $\bar{E}_j$ denote
respectively the velocity, momentum and energy of $\nu_j$ when
$\bar{E}_j=M_\pi-M_i$, which corresponds to the kinematics of $\pi\rightarrow
l_i\nu_j$ with $\pi,l_i$ at rest. The detailed calculation of
Eq.\,\eqref{Psi:x,t:approx} is carried out in appendix
\ref{ap:Psi:x,t:approx}.

We can verify the normalization of Eq.\,\eqref{Psi:x,t:approx}:
\eq{
\label{norm:1}
\sum_{rs}\int\Dp{x}|\Psi^{rs}_{ij}(\bx,t)|^2=
\frac{\gamma}{\Gamma_\pi}|U_{ij}|^2\tilde{\Gamma}_j(\bar{E}_j)
[1+4\frac{\Gamma_\pi}{\bar{p}_j} e^{-\Gamma_\pi t}\sin(\bar{p}_j\bar{v}_jt)]
\,.
}
Equation \eqref{norm:1} guarantees that
\eq{
\sum_{ij,rs}\int\Dp{x}|\Psi^{rs}_{ij}(\bx,t)|^2=1\,,
}
for $t\gg 1/\Gamma_\pi$ or $\frac{\Gamma_\pi}{\bar{p}_j}\ll 1$.
Therefore, the total probability is conserved by the approximation.
One can not, however, recover the flavor oscillation phenomenon from the 
approximate wave function \eqref{Psi:x,t:approx} because the phase is common to
all states $l_i\nu_j$ with fixed $i$.
Numerical studies can be performed to test the accuracy of the approximation,
although appropriate care should be taken with the enormous difference between
the scales of $m_j$ and $M_\pi,M_i$.

To visualize the localization aspects of the neutrinos created, we can
calculate the radial probability density for the creation of neutrinos $\nu_j$
at radius $r$ and time $t$ jointly with $l_i$,
\eqarr{
\rho_{ij}(r,t)&=&
\sum_{rs}\int d\Omega_{\bx}\,r^2|\Psi^{rs}_{ij}(\bx,t)|
\,,\\&=&
\gamma|U_{ij}|^2\tilde{\Gamma}_j(M_\pi-M_i)
\frac{1}{\bar{v}_j}
\big[
\theta(\tau_-)e^{-\Gamma_\pi\tau_-}
\cr&&~~+
\theta(\tau_+)e^{-\Gamma_\pi\tau_+}
+2\theta(\tau_+)\theta(\tau_-)\cos(\bar{p}_j r)e^{-\Gamma_\pi t}
\big]
\,.
}
For $t\gg 1/\Gamma_\pi$, we see only the first term containing
$\tau_-$ contributes. Thus $\rho_{ij}(r,t)$ has a triangular shape with an
abrupt peak in $r=\bar{v}_jt$ and a exponential tail for $r<\bar{v}_jt$, being
negligible for $r>\bar{v}_jt$.
The neutrino is then roughly localized in the region
$\bar{v}_j(t-1/\Gamma_\pi)\lesssim r\le \bar{v}_jt$. The size of the wave packet is
roughly $\bar{v}_j/\Gamma_\pi$ as is usually assumed in rough
estimates\,\cite{akhmedov,kayser:81}.

The most faithful state that we can construct to describe the neutrino flavor
states created from pion decay jointly with a charged lepton $l_i$ with
momentum $\bq$ and spin $r$ is
\eqarr{
\label{nuj:flavor}
e^{iM_it}{}_0\braket{l_i(\bq,r)}{\pi(t)}&=&
\sum_{j,s}\int\Dp{p}\psi_\pi(\bq+\bp)
e^{-iE_j(\bp)t}\ket{\nu_{j}(\bp,s)}_0
\cr&&~\times
\frac{gU_{ij}\eta^{rs}_j(\bp)}
{h(M_i+E_j(\bp)+i\epsilon)}
\Big[
1-e^{-i(M_\pi-M_i-E_j)t}e^{-\Gamma_\pi t/2}
\Big]
\,
}
where $\psi_\pi$ is the function appearing in the definitions
\eqref{pi:0} and \eqref{linuj:0}, corresponding to the pion momentum wave
function. In the limit $t\gg \Gamma_\pi$, we can approximate $m_j\rightarrow 0$
in all factors except in the exponent since $m_j\ll |\bp|,M_i,M_\pi$.
Equation\,\eqref{nuj:flavor} becomes
\eq{
\label{nuj:flavor:ap}
e^{iM_it}{}_0\braket{l_i(\bq,r)}{\pi(t)}\approx 
\sum_{j,s}\int\Dp{p}\psi_\pi(\bq+\bp)
e^{-iE_j(\bp)t}U_{ij}\ket{\nu_{j}(\bp,s)}_0
\frac{g\eta^{rs}(\bp)}
{h(M_i+|\bp|+i\epsilon)}
\,.
}
We could then write, for $t=0$,
\eq{
{}_0\braket{l_\alpha(\bq,r)}{\pi(t)}\approx 
\sum_{s}\int\Dp{p}\psi_\pi(\bq+\bp)
\ket{\nu_{\alpha}(\bp,s)}_0
\frac{g\eta^{rs}(\bp)}
{h(M_\alpha+|\bp|+i\epsilon)}
\,,
}
where $l_i=l_\alpha$, $M_i=M_\alpha$ and 
\eq{
\ket{\nu_{\alpha}(\bp,s)}_0\equiv\sum_{j} U_{\alpha j}\ket{\nu_j(\bp,s)}_0
\,.
}
If we impose $\psi_\pi(\bp)=\delta^3(\bp)$ we obtain exactly $\bq=-\bp$ for the
momenta of the charged leptons ($\bq$) and neutrinos ($\bp$). In general, 
the presence of the function $\psi_\pi$ ensures momentum conservation and
forces the amplitude \eqref{nuj:flavor} to be appreciable only around
$\bq=-\bp$ where $\bp$ has magnitude satisfying $E_j(\bp)\approx M_\pi-M_i$.
Equation \eqref{nuj:flavor} also clarifies the contribution of the intrinsic
momentum uncertainty $\Delta p$ (intrinsic of $\psi_\pi$) of the parent particle
that is inherited by the daughter particles, apart from the contribution of the
decay width $\Gamma_\pi$: the smallest between $\Gamma_\pi$ and $\Delta p$
dominates.
Other transition amplitudes of $\ket{\pi(t)}$ with respect to the free states
$\ket{\pi}_0$ and $\ket{l_i(r),\nu_j(\bp,s)}_0$ can be found in appendix
\ref{ap:WW}. One can identify some similarity with the Wigner-Weisskopf
approximation used in systems with couplings between discrete and
continuum energy levels.

\section{Discussions}
\label{sec:discussion}

One knows neutrino flavor oscillation ceases when quantum coherence is lost due
to the lack of spatial overlap among the mass eigenstates that compounds the
flavor state neutrinos $\nu_\alpha$\,\cite{giunti:cease,giunti:coh}. The
characteristic time (distance) scale for such phenomenon to occur is usually
very large for neutrinos and is given by $\delta x/\Delta v_{ij}$, where $\delta
x$ is the characteristic spatial size of the neutrino wave packets and $\Delta
v_{ij}$ is the velocity difference of the $\nu_i$ and $\nu_j$ that compound the
flavor state. In this model, in the pion restframe,
\eq{
\delta x_j\approx\bar{v}_j/\Gamma_\pi\,.
}
Neutrino flavor oscillation occurs because the neutrinos produced in channels
$\pi\rightarrow l_2\nu_1$ and $\pi\rightarrow l_2\nu_2$ interfere coherently
and remain interfering in the same region of space as long as $t\ll \delta
x_j/\Delta v_{ij}$.
But what prevents the channels $\pi\rightarrow l_1\nu_1$ and
$\pi\rightarrow l_2\nu_1$ from interfering?
For real pion decays such interference is not observable even in principle
since these two channels have very different probabilities to occur because of
helicity suppression in weak decays. In this model, however, the
branching ratios are comparable and may mimic other real decays.

One should notice that the neutrinos created in channels $\pi\rightarrow
l_1\nu_1$ and $\pi\rightarrow l_2\nu_1$, for example, may have significant
spatial overlap.
We see that $\Delta_Mv(m_j)$ defined as
\eq{
\Delta_M v(m_1)\equiv \bar{v}_1(M_1)-\bar{v}_1(M_2)\,,
}
may be of order of
\eq{
\Delta_m v(M_i)\equiv \bar{v}_1(M_i)-\bar{v}_2(M_i)\,,~~
}
unless the neutrino masses are nearly degenerate $\Delta
m_{ij}^2\ll m_i^2, m^2_j$. 
We use the notation $\bar{v}_j(M_i)$ instead of $\bar{v}_j$ throughout 
this discussion to avoid ambiguities.
It means that the non interference of these channels is probably due to the
charged leptons, but its detailed account deserves further study.
Simple calculations show\,\cite{akhmedov} that the big mass difference between
charged leptons make them lose spatial coherence over a distance of atomic
length or less for most decaying processes. Is this lost of spatial
coherence responsible for the incoherence of the channels with definite flavor 
$l_1=e$ or $l_2=\mu$? Another possibility is that the detection of charged
leptons is what makes the coherent superposition of the four channels
$\pi\rightarrow l_i\nu_j$, $i,j=1,2$, reduce into two incoherent $\pi\rightarrow
l_1\nu$ and $\pi\rightarrow l_2\nu$ quantum processes.
These two explanations are, however, very distinct since the former is
detection independent while the latter is detection dependent.

Concerning the absence of the interference between the channels $\pi\rightarrow
l_1\nu$ and $\pi\rightarrow l_2\nu$, there is one ingredient that could play
an important role, i.e., the spatial entanglement. This question, however, can
not be discussed within this model due to the lack of spatial entanglement
between $l_i$, in the static limit, and $\nu_j$.

Another known situation when neutrino oscillations can be suppressed is the case when
it is possible to probe the neutrino kinematical quantities with enough precision to
single out one neutrino mass eigenstate that is being created in the
process\,\cite{kayser:81}. In the language of this article it means all the four pion
decay channels proceeds incoherently. Such suppression, of course, can be induced by
localization properties of the source or detector that produces and detects the
neutrinos\,\cite{kiersWeiss}. Conditions necessary for neutrino oscillation can be
analyzed, for instance, within a realistic QFT description\,\cite{grimus}.
In that case, the condition for neutrino oscillation requires 
[Eq.\,(4.7) of Ref.\,\cite{grimus}] 
\eq{
\label{grimus}
\frac{\Delta m^2}{2E_\nu}\lesssim \Gamma
\,,
}
if the intrinsic momentum spread of the decaying particle, apart from the decay
width $\Gamma$, is small compared to its mass.
Condition \eqref{grimus} is exactly the condition necessary to prevent one from
knowing which neutrino mass eigenstate is being created\,\cite{kayser:81}, when the
momentum uncertainty is of the order of $\Gamma$, which is also the case of this
work. 
Notice, however, that in this work we do not treat the propagation of neutrinos but
rather focus the attention on the creation process.
Surprisingly, condition \eqref{grimus}, with $\ll$ instead of $\lesssim$, is exactly
what is needed to guarantee, initially, approximate neutrino flavor definition. Such
analysis involves estimating the first flavor violating term of
Eq.\,\eqref{creationP:2} that follows after Eq.\,\eqref{pi->l,Unu}.

Therefore, the study of this Lee-type model sheds light into the question of what is
a superposition state of definite flavor. We could confirm, within this model, the
result that once the channels with charged leptons with definite masses are singled
out, coherent creation of neutrino flavor states is allowed because the kinematical
contributions from each neutrino mass eigenstate to their creation probabilities are
negligible. Therefore the creation probability amplitudes of neutrinos are mostly
independent of neutrino masses (mass differences to be precise), enabling the
decoupling of Eq.\,\eqref{pi->l,Unu} into the creation probability times the flavor
conversion probability. The conversion probability, however, depends crucially on the
tiny neutrino mass differences. The neutrino flavor states $\nu_\alpha$ can be
created jointly with the charged lepton $l_i=l_\alpha$ and the wrong flavor is
initially absent. On the other hand, if we try to calculate, as in
Eq.\,\eqref{pi->Ul,nu}, the creation probability of the coherent superposition of
charged leptons jointly with the neutrino $\nu_1$, for instance, there is no meaning
to attribute the name ``flavor'' for that superposition of charged leptons because
such flavor is not created in an unambiguous manner but both the correct and the
wrong flavors would be created initially.

To be extremely precise, the notion of neutrino flavor states 
rely on the approximation of negligible neutrino masses,
as could be seen in Eq.\,\eqref{pi->l,Unu}. For that reason, there is a
negligible probability to create the wrong flavor $\nu_e$, for example, jointly
with the charged lepton $l_2=\mu$. Such probability is calculable
but negligible because of the proportionality to the neutrino masses. Such
flavor indefiniteness due to the neutrino mass differences can be also seen in
other attempts to exactly define what is a neutrino flavor state such as in
Refs.\,\onlinecite{BV} or \onlinecite{giunti:qft}. The exact definition is
always blurred by the differences in the kinematical or dynamical factors that
depend, strictly speaking, on the neutrino masses, that in turn, can not be
factored out as a common factor. For example, the spinorial character of
neutrinos prevent even a description as simple as a mixed free second quantized
theory to provide exact initial neutrino flavor definition\,\cite{ccn:no12}.
Another simple treatment that leads to initial flavor violation is the adoption
of general kinematical conditions when scalar wave packets are
used\,\cite{ccn:WP}.

To summarize, we could confirm two aspects, within a Lee-type model: (a) 
the spatial size of the neutrinos produced through a decaying particle with
decay rate $\Gamma$ is $v/\Gamma$ for the parent particle at rest. This relation
is commonly used in estimates of the neutrino wave packet sizes but this model
provides an explicit example satisfying such relation. (b) The coherent
creation of neutrino flavor states is possible because the kinematical
contribution of neutrinos to their creation probabilities are negligible,
allowing to define the flavor state as a superposition of mass eigenstates in
the expected way. We should remark that the second conclusion is, in principle,
applicable to more realistic decays because the main ingredient is universal:
neutrino mass differences, compared to the charged lepton mass differences, 
contribute negligibly to their creation probabilities.

\appendix
\section{The function $\phi_0$}
\label{ap:phi0}

The function $\phi_0$ appearing in Eq.\,\eqref{phi0:def} can be written in the
following simple form:
\eqarr{
\phi_0(E_\pi)&=&
2\sum_{ij}\int\Dp{p}\frac{\tilde{f}^2_j(\bp)|U_{ij}|^2}
{M_i+E_j(\bp)-E_\pi}
\cr&=&\label{phi0:2}
\frac{g^2_0}{2\pi^2}\sum_{ij}
\int_0^{\infty}dp\,\frac{p^4f^2(p)}{E_j(E_j+m_j)}
\frac{|U_{ij}|^2}{M_i+E_j-E_\pi}
\\&=&\label{phi0:3}
\gamma_0\sum_{ij}
\int_{m_j}^{\infty}dE\,\theta(E-m_j)\sqrt{E^2-m^2_j}(E-m_j)f^2(E)
\frac{|U_{ij}|^2}{M_i+E-E_\pi}
\\&=&\label{phi0:4}
\gamma_0\sum_{ij}|U_{ij}|^2\tilde{\phi}_j(E_\pi-M_i)
\,.
}
In Eq.\,\eqref{phi0:2} we have assumed the cutoff function is isotropic
$f(\bp)=f(|\bp|)$, $|\bp|=p$. In addition, it is assumed in Eq.\,\eqref{phi0:3}
that $f$ is a function of $E$ only, i.e., $f(\bp)\rightarrow f(E_j(\bp))$.
We used Eqs.\,\eqref{g0}, \eqref{phitilde} and
\eq{
\eta^{rs}_j(\bp)=\frac{p}{\fpi}\frac{f(|\bp|)}{\sqrt{2E_j}\sqrt{E_j+m_j}}
(-i\eta_C)(\hat{\bp}\sp\bs{\sigma}\sigma_2)_{rs}
}
\eq{
\sum_{rs}|\eta^{rs}_j(\bp)|^2=
2\big(\tilde{f}_j(\bp)\big)^2\,,
}
where
\eq{
\tilde{f}_j(\bp)=
\frac{1}{\fpi}
\frac{f(\bp)}{\sqrt{2E_j}}
\,.
}

If we extend the variables to small complex values, we obtain
\eq{
\tilde{\phi}_j(x\pm i\epsilon)=
\tilde{\phi}_j(x)\pm\frac{i}{2}\tilde{\Gamma}_j(x)
\,,
}
where
\eq{
\tilde{\Gamma}_j(x)\equiv
2\pi\theta(x-m_j)(x-m_j)\sqrt{x^2-m^2_j}f^2(x)
\,.
}
We see $\tilde{\Gamma}_j(x)\ge 0$ for all $x$ while $\tilde{\Gamma}'_j(x)\ge 0$
for $x>m_j$ and $x$ much lower than the cutoff scale of $f$, assuming $f$ is
a smooth function on that region.
\section{Completeness of the states $\ket{l_i(r),\nu_j(\bp,s)}^{\oi}$}
\label{linuj:completo}

We prove here the states $\ket{l_i(r),\nu_j(\bp,s)}^{\oi}$ in
Eq.\,\eqref{linuj:d} complete the Hilbert space in the sector of one $\pi$ or
$l_i\nu_j$. We calculate
\eqarr{
\sum_{ij}\int\Dp{p}\ket{l_i,\nu_j(\bp)}
\bra{l_i,\nu_j(\bp)}
&=&
C\ket{\pi}_0{}_0\bra{\pi}
+
\sum_{ij}\int\Dp{p}\ket{l_{i},\nu_{j}(\bp)}_0
{}_0\bra{l_{i},\nu_j(\bp)}
\cr&&
+\sum_{ij,i'j'}\int\Dp{q}\Dp{q'}
\ket{l_{i'},\nu_{j'}(\bq)}_0
A_{i'j',ij}(\bq,\bq')
{}_0\bra{l_{i},\nu_j(\bq')}
\cr&&
+
\sum_{ij}\int\Dp{q}
B_{ij}(\bq)\ket{l_i,\nu_j(\bq)}_0
{}_0\bra{\pi}
+ h.c.
}
The spin indices are suppressed.
We get
\eqarr{
C&=&
Z_2\sum_{ij}\int\Dp{p}|\beta_{ij}(\bp)|^2\,,
\\
A_{i'j',i''j''}(\bq,\bq')&=&
\Big[
\alpha_{i''j'',i'j'}(\bq',\bq)
+\alpha^*_{i'j',i''j''}(\bq,\bq')
+\sum_{ij}\int\Dp{p}\alpha_{ij,i'j'}(\bp,\bq)
\alpha^*_{ij,i''j''}(\bp,\bq')
\Big]
\,,~~~~\\
B_{i'j'}(\bq)&=&
Z_2^{\meio}\beta^*_{i'j'}(\bq)
+\sum_{ij}\int\Dp{p}Z_2^{\meio}\beta^*_{ij}(\bp)\alpha_{ij,i'j'}(\bp,\bq)
\,.
}

The first coefficient yields
\eqarr{
C&=&
Z_2\int\Dp{p}|\beta_{ij}(\bp)|^2
\\&=&
\sum_{ij,rs}\int\Dp{p}
\frac{|g_0U_{ij}\eta^{rs}_j(\bp)|^2}{|h_0(M_i+E_j(\bp)+i\epsilon)|^2}
\\&=&
\frac{\gamma_0}{\pi}\int_{M_1+m_1}^{\infty} dE\sum_{ij}
\frac{\tilde{\Gamma}_j(E-M_i)|U_{ij}|^2}{|h_0(E+i\epsilon)|^2}
\\&=&
\frac{\rm Im}{\pi}
\int_{M_1+m_1}^{\infty}dE\frac{h_0(E+i\epsilon)}{|h_0(E+i\epsilon)|^2}
\\&=&
-\frac{1}{2\pi i}
\int_{M_1+m_1}^{\infty}dE
\Big[\frac{1}{h_0(E+i\epsilon)}-\frac{1}{h_0(E-i\epsilon)}\Big]
\\&=&
-\frac{1}{2\pi i}
\int_{P} \frac{dz}{h_0(z)}
\\&=&
1
\,.
}
The contour $P$ is a path along the real axis coming from $\infty$ to $m_1+M_1$
below the real axis then going from $m_1+M_1$ to $\infty$ over the real axis. If
the contour is closed with the aid of a very large circle, the integral over the
closed contour is zero due to the absence of poles. The integral over the circle
is equal to 1 due to $h_0(z)\sim z$.

We obtain $A_{i'j',i''j''}(\bq,\bq')=0$ from
\eqarr{
\sum_{ij}\int\Dp{p}\alpha_{ij,i'j'}(\bp,\bq)\alpha^*_{ij,i''j''}(\bp,\bq')
&=&
\sum_{ij}\int\Dp{p}
\frac{Z_2|\beta_{ij}(\bp)|^2g^2_0\eta_{j'}(\bq)\eta^*_{j''}(\bq')
U_{i'j'}U^*_{i''j''}}
{M_i+E_j(\bp)-M_{i'}-E_{j'}(\bq)+i\epsilon}
\cr&&\hs{3.3em}\times
\frac{1}{M_i+E_j(\bp)-M_{i''}-E_{j''}(\bq)-i\epsilon}~~
\\&=&
g^2_0\eta_j(\bq)\eta^*_j(\bq')U_{i'j'}U^*_{i''j''}
\frac{(-1)}{2\pi i}
\int_P \frac{dz}{h_0(z)}
\frac{1}
{z-M_{i'}-E_{j'}(\bq)+i\epsilon}
\cr&&\hs{3.3em}\times
\frac{1}{z-M_{i''}-E_{j''}(\bq)-i\epsilon}~~
\,,
\\&=&
\frac{
-g^2_0\eta_j(\bq)\eta^*_j(\bq')U_{i'j'}U^*_{i''j''}
}{
(M_{i'}+E_{j'}(\bq)-M_{i''}-E_{j''}(\bq')-i\epsilon)
}
\cr&&\times
\Big[
\frac{1}{h_0(M_{i'}+E_{j'}(\bq)-i\epsilon)
}
-
\frac{1}{h_0(M_{i''}+E_{j''}(\bq)+i\epsilon)
}
\Big]
\\&=&
-\alpha^*_{i'j',i''j''}(\bq,\bq')
-\alpha_{i''j'',i'j'}(\bq',\bq)
\,.
}
Analogously one can prove 
\eq{
B_{ij}(\bq)=0 \,.
}
\section{Calculation of Eq.\,$\text{\eqref{Psi:x,t:approx}}$}
\label{ap:Psi:x,t:approx}

The gap between Eqs.\,\eqref{Psi:x,t} and \eqref{Psi:x,t:approx} is filled by
\eqarr{
\Psi^{rs}_{ij}(\bx,t)
&=&
\label{Psi:1}
\frac{gU_{ij}}{\fpi}
\int_0^{\infty}dp\,\eta^{rs}_j(p\hat{\bx})
\frac{4\pi}{i}\frac{\partial}{\partial r}\Big(\frac{\sin(pr)}{r}\Big)
\frac{e^{-i(M_i+E_j(\bp))t}}{h(E_j+M_i+i\epsilon)}
\,,\\&\approx&
\label{Psi:2}
\frac{4\pi}{ir}\frac{gU_{ij}}{\fpi}
\int_{m_j+M_i}^{\infty} \frac{dE}{v_j}
\eta^{rs}_j(p_j\hat{\bx})\cos p_jr
\frac{e^{-iEt}}{h(E)+\frac{i}{2}\Gamma(E)}
\,,\\&\approx &
\label{Psi:3}
\frac{2\sqrt{\pi}}{ir}
\gamma^{\meio}U_{ij}
\int_{m_j+M_i}^{\infty} \frac{dE}{v_j}
\eta^{rs}_j(p_j\hat{\bx})\cos p_jr
\frac{e^{-iEt}}{E-M_\pi +\frac{i}{2}\Gamma_\pi}
\,,\\&\approx &
\label{Psi:4}
\frac{2\sqrt{\pi}}{ir}
\gamma^{\meio}U_{ij}
\eta^{rs}_j(\bar{p}_j\hat{\bx})\frac{e^{-iM_\pi t}}{\bar{v}_j}
\int_{-\infty}^{\infty} dE
\cos\big((\bar{p}_j+\frac{E}{\bar{v}_j})r\big)
\frac{e^{-iEt}}{E +\frac{i}{2}\Gamma_\pi}
\,.
}
To get to Eq.\,\eqref{Psi:1} we made use of
\eq{
\int\Dp{p}\eta^{rs}_j(\bp)e^{i\bp\sp\bx}
=
\int_0^{\infty}dp\,\eta^{rs}_j(p\hat{\bx})
\frac{4\pi}{i}\frac{\partial}{\partial r}\Big(\frac{\sin(pr)}{r}\Big)
\,.
}
To get to Eq.\,\eqref{Psi:2} we neglected the nonradiative term containing
$r^{-2}$ and used the change of variables $p=|\bp|\rightarrow
E=E_j(\bp)+M_i$. We also used the shorthands $p_j\equiv
p_j(E_j)=\sqrt{E_j^2-m^2_j}$, $v_j\equiv v_j(E_j)=\frac{p_j(E_j)}{E_j}$
for $E_j=E-M_i$.
To get to Eq.\,\eqref{Psi:3} we approximate
\eq{
\label{h:approx}
h(E+i\epsilon)\approx E-M_\pi + \frac{i}{2}\Gamma_\pi\,,
}
by Taylor expanding around $E=M_\pi$. Notice that $h(M_\pi)=0$ and
$h'(M_\pi)=1$. The use of the approximation \eqref{h:approx} requires 
\eq{
|\Gamma'(M_\pi)|\ll 1, ~~
|h''(M_\pi)\Gamma_\pi|\ll 1\,.
}

In Eq.\,\eqref{Psi:4} we approximated $E=M_\pi-M_i$ in the whole integrand,
except in the exponentials and in the term $h^{-1}$. It is assumed that
these functions vary slowly in the interval $E\approx M_\pi\pm \Gamma_\pi$. 
For $p_j$ in the cosine we approximated
\eq{
p_j\approx \bar{p}_j+\frac{E}{\bar{v}_j}\,,
}
where $E$ is obtained from the shift $E\rightarrow E+M_\pi$ compared to
Eq.\,\eqref{Psi:3}. It is also assumed that
\eq{
M_\pi-\Gamma_\pi \gtrsim m_j+M_i\,,
}
which allows us to replace the lower limit of the integral by $-\infty$ without
changing the integral appreciably.
At last, Eq.\,\eqref{Psi:x,t:approx} is recovered by using the integral
\eq{
\int_{-\infty}^{\infty}dE\,\frac{e^{-iEt}}{E+i\lambda}=
-2\pi i\theta(t)e^{-\lambda t}
\,,~~\lambda>0\,,
}
computed in the complex plane closing the contour on the upper or lower half
plane.

\section{Approximate eigenstate}
\label{ap:pi:approx}

We can see the state in Eq.\,\eqref{pi:lambda:0} can not be an exact eigenstate
of the total Hamiltonian $H$ by calculating the eigenvalue equation, analogous
to Eq.\,\eqref{h0=0}, with the complex eigenvalue
\eq{
E_\pi=M_\pi+i\lambda\,.
}
We thus obtain
\eq{
M_\pi+i\lambda+\phi_0(M_\pi+i\lambda)=0\,.
}
The real part is just Eq.\,\eqref{h0=0}.
The imaginary part yields
\eq{
|\lambda|= -\frac{1}{2}\frac{\Gamma_0(M_\pi)}{1+\phi'_0(M_\pi)}=
-\frac{\Gamma_\pi}{2}\,,
}
which can never be satisfied because of the minus sign.
Therefore if both $|\lambda|$ and $\Gamma_\pi$ are small, the state
\eqref{pi:lambda:0} represents an approximate eigenstate of $H$.

\section{Wigner-Weisskopf approximation}
\label{ap:WW}

We can compare the results of Eq.\,\eqref{e-Gt} with the transition
probabilities with respect to the free (bare) states. Such transition amplitudes
resemble the Wigner-Weisskopf (WW) approximation\,\cite{ww} employed in atomic
physics.

For the transition amplitude of the approximate pion state to the free pion we
obtain
\eqarr{
Z_2^{-\meio}{}_0\braket{\pi}{\pi(t)}&=&
\braket{\pi(0)}{\pi(t)}
\\&\approx&
\label{pi->pi}
e^{-i(M_\pi-i\Gamma_\pi/2)t}\,,
}
where the factor $Z_2^{-\meio}$ is necessary from renormalization.

For the transition amplitude of the approximate pion state to the free
$l_i\nu_j$ states we obtain
\eqarr{
{}_0\braket{l_i(r),\nu_j(\bp,s)}{\pi(t)}&=&
C^{rs}_{ij}(\bp)\Big[e^{-i(M_i+E_j)t}
\cr&&~~
+~h(M_i+E_j+i\epsilon)\sum_{r's',i'j'}
\int\Dp{p'}
\frac{|C^{r's'}_{ij}(\bp')|^2e^{-i(M_{i'}+E_{j'})t}}
{M_{i'}+E_{j'}-M_i-E_j-i\epsilon}
\Big]
\\&\approx&
\label{pi->li,nuj}
C^{rs}_{ij}(\bp)e^{-i(M_i+E_j)t}
\Big[1-e^{-i(M_\pi-M_i-E_j)t}e^{-\frac{\Gamma_\pi}{2}t}\Big]\,.
}
Equations \eqref{pi->pi} and \eqref{pi->li,nuj} illustrate clearly the
irreversible probability flow from the free state $\ket{\pi}_0Z^{-\meio}$ to
the free states $\ket{l_i(r),\nu_j(\bp,s)}_0$.

\acknowledgments
This work was partially supported by {\em Fundação de Amparo à Pesquisa do
Estado de São Paulo} (Fapesp) and 
{\em Conselho Nacional de Desenvolvimento Científico e Tecnológico} (CNPq).



\begin{thebibliography}{99}
\bibitem{endnote0}
The vectors $e_r$ are the two-dimensional canonical column vectors defined by
$(e_r)_a=\delta_{ar}$.
\bibitem{endnote1}
This is another distinctive feature of the model which renders it exactly
solvable.
\end{thebibliography}
\end{document}